# A Comprehensive Study of Groundbreaking Machine Learning Research: Analyzing Highly Cited and Impactful Publications across Six Decades

Absalom E. Ezugwu[1,*], Japie Greeff[2], Yuh-Shan Ho[3,*],

[1]Unit for Data Science and Computing, North-West University, 11 Hoffman Street, Potchefstroom, 2520, South Africa

[2]School of Computer Science and Information Systems, Faculty of Natural and Agricultural Sciences, North-West University, Vanderbijlpark, South Africa

[3]Trend Research Centre, Asia University, No. 500, Lioufeng Road, Taichung 41354, Taiwan

*Corresponding Authors (Absalom.ezugwu@nwu.ac.za; ysho@asia.edu.tw)

**Abstract**. Machine learning (ML) has emerged as a prominent field of research in computer science and other related fields, thereby driving advancements in other domains of interest. As the field continues to evolve, it is crucial to understand the landscape of highly cited publications to identify key trends, influential authors, and significant contributions made thus far. In this paper, we present a comprehensive bibliometric analysis of highly cited ML publications. We collected a dataset consisting of the top-cited papers from reputable ML conferences and journals, covering a period of several years from 1959 to 2022. We employed various bibliometric techniques to analyze the data, including citation analysis, co-authorship analysis, keyword analysis, and publication trends. Our findings reveal the most influential papers, highly cited authors, and collaborative networks within the machine learning community. We identify popular research themes and uncover emerging topics that have recently gained significant attention. Furthermore, we examine the geographical distribution of highly cited publications, highlighting the dominance of certain countries in ML research. By shedding light on the landscape of highly cited ML publications, our study provides valuable insights for researchers, policymakers, and practitioners seeking to understand the key developments and trends in this rapidly evolving field.

## 1. Introduction

Machine learning (ML) has undergone a transformative evolution within the field of artificial intelligence, bringing about significant changes across numerous industries and scientific domains (Ahmed, Jeon & Piccialli, 2022; Zhong et al., 2021; Ezugwu et al., 2023; Ezugwu et al., 2020; Yao et al., 2023; Kreuzberger, Kühl, and Hirschl, 2023; Masini et al., 2023). The rapid progress of ML techniques and algorithms has resulted in a proliferation of research publications in this field. Consequently, identifying and analyzing influential and highly cited publications have become crucial amidst the extensive literature available. These publications have made significant contributions to the advancement of ML research and its application in various domains (Kotsiantis, Zaharakis & Pintelas, 2006; Kotsiantis, Zaharakis & Pintelas, 2007; Mahesh, 2022).

This paper probes into the domain of bibliometric exploration in order to provide insights into the landscape of highly cited and influential publications in the field of ML research. Through the application of bibliometric analysis, our objective is to uncover crucial trends, influential authors, top journals, and significant themes within this dynamic field. This investigation serves not only to offer valuable insights into the progress of ML research but also to assist researchers, practitioners, and decision-makers in identifying seminal works and gaining a deeper understanding of the field's direction. Numerous studies have previously presented bibliometric analyses focusing on specific research areas within machine learning. For instance, De Felice and Polimeni (2020) conducted a study on disease forecasting, while Kim, Lee, and Park (2021) explored the application of ML in mental health research. Yu, Xu, and Wang (2021) investigated research trends in support vector machines, and Su, Peng, and Li (2021) examined ML research trends in engineering. These studies exemplify the diverse range of applications within ML that have been subject to bibliometric analysis.

Utilizing bibliometric analysis provides a systematic and quantitative means to evaluate the impact and importance of academic publications. This approach involves examining citation patterns, co-authorship networks, and publication trends (Rosas et al., 2011; Glänzel & Schubert, 2005; Ezugwu et al., 2021a; Ezugwu et al., 2021b). By employing these techniques, we gain the ability to uncover the most influential research papers, influential authors, and collaborations that have played a crucial role in shaping the field of ML research from various perspectives. Furthermore, this analysis allows us to identify the prominent research areas, methodologies, and applications that have garnered significant attention and citation within the scholarly community. More so, understanding the landscape of highly cited and high impact ML publications is vital for various stakeholders. Researchers can gain a comprehensive overview of the field, identify potential research gaps, and discover fruitful directions for their investigations (Waheed et al., 2018; Riahi et al., 2021). Funding agencies and policymakers can leverage these insights to allocate resources strategically and promote impactful research. Moreover, practitioners and industry professionals can benefit from this analysis by staying abreast of the latest advancements, prominent researchers, and cutting-edge methodologies in the field.

In this article, we present our findings from a meticulous bibliometric analysis of ML research, focusing on highly cited and high impact publications. Through this exploration, we aim to provide a comprehensive understanding of the scholarly impact in the field of ML, offering a roadmap for future research directions and highlighting the most influential contributions. Our study serves as a valuable resource for researchers, practitioners, and stakeholders in the ML community, facilitating informed decision-making and fostering further advancements in this rapidly evolving field.

The remaining sections of this paper are structured as follows: In Section 2, we outline the methods utilized for data gathering and the various search strategies employed for conducting the proposed bibliometric analysis study. Section 3 provides a comprehensive presentation of the results and discussion, including a detailed description of the bibliometric study and an analysis of the data collection pertaining to different ML research publications. Additionally, this section offers an overview of the discussion surrounding the top cited ML publications. Finally, in Section 4, we draw conclusions based on the findings of this study.

## 2. Data Collection and Search Strategy

In this study, we employed comprehensive data collection techniques to gather ML research publications from various sources. These sources include leading academic journals, conference proceedings, and other relevant scholarly outlets. By utilizing a diverse range of search strategies, we ensured the inclusiveness and representativeness of the collected dataset.

Moreover, for this study, we retrieved data from the Clarivate Analytics Web of Science Core Collection, specifically the online version of the Science Citation Index Expanded (SCI-EXPANDED). The dataset used in our analysis was updated on 18 May 2023, ensuring the inclusion of the most recent publications in the field. To construct a comprehensive dataset, we employed a carefully designed search strategy. We utilized quotation marks (" ") and the Boolean operator "or" to ensure that at least one search keyword appeared in terms of TOPIC (title, abstract, author keywords, and Keywords Plus). The search encompassed the period from 1900 to 2022, spanning over a century of ML literature.

Our search was primarily focused on the keyword "machine learning." However, to account for variations in terminology and to capture a broader range of relevant publications, we included additional terms such as "machine learned," "machine learn," "machine learners," "machines learning," "machining learning," "machine learner," "machine learnings," "machines learn," "machine learns," "machine learnable," "maching learning," "learning of machine," and "machin learning." Furthermore, we incorporated misspelled terms like "machine learnt" and "machine learnig" to account for potential typographical errors. Additionally, terms lacking spaces, such as "machine learningbased," "machine learningmethods," "machine learnin," "machine learningalgorithm," "machine learningclassifiers," and "machine learningmetrics," were included to capture relevant documents within the SCI-EXPANDED database. By adopting this comprehensive search approach, we aimed to ensure that our analysis results are as accurate and inclusive as possible, encompassing a wide range of documents related to the field of ML research.

**2.1 Assessing publication impact**

To gauge the impact of publications in this study, we employed several citation indicators derived from the Web of Science Core Collection. These indicators provide valuable insights into the citation performance of individual publications. The following citation indicators were utilized:

- $C_{\text{year}}$: This indicator represents the number of citations a publication received from the Web of Science Core Collection in a specific year. For example, $C_{2022}$ denotes the citation count in the year 2022 (Ho, 2012).

- $TC_{\text{year}}$: The $TC_{\text{year}}$ reflects the total number of citations a publication has received from the Web of Science Core Collection since its publication year up until the end of the most recent year (2022 in our study, $TC_{2022}$) (Wang et al., 2011).

- $CPP_{\text{year}}$: The $CPP_{\text{year}}$ stands for the average number of citations per publication within a particular year. Specifically, $CPP_{2022}$ is calculated as $TC_{2022}$ divided by $TP$, which represents the total number of publications (Ho, 2013).

The use of these citation indicators, namely $C_{\text{year}}$, $TC_{\text{year}}$, $CPP_{\text{year}}$, offers distinct advantages. These indicators ensure consistency and repeatability in our analysis, compared to directly using the number of citations from the Web of Science Core Collection (Ho and Hartley, 2016a). To identify highly cited publications, we employed a criterion where publications with $TC_{2022}$ of 100 or more were selected. This threshold allows us to focus on publications that have received substantial attention and recognition within the focused field (Ho, 2014a). By utilizing these citation indicators and criteria, we aim to identify and highlight the most influential and highly cited publications in the field of ML research. These indicators provide a quantitative measure of the impact and visibility of individual publications, offering valuable insights into the scholarly contributions that have significantly shaped the field of ML.

**2.2 Methodology and data analysis**

In this study, we employed a meticulous approach to identify and analyze highly cited ML documents. A total of 5,402 documents with a $TC$ (total citations from the Web of Science Core Collection) of 100 or more were searched and retrieved from SCI-EXPANDED, covering the period from 1959 to 2022. The data used in our analysis were updated as of 18 May 2023. To conduct our analysis, we downloaded the full records of these highly cited documents from SCI-EXPANDED, along with the number of citations received in each year for each document. The downloaded data were then imported into Microsoft 365 Excel for further analysis. Manual coding was performed to enhance the data and extract relevant information (Li and Ho, 2008; Al-Moraissi et al., 2023).

Various functions available in Microsoft 365 Excel, such as Counta, Concatenate, Filter, Match, Vlookup, Proper, Rank, Replace, Freeze Panes, Sort, Sum, and Len, were utilized to process and analyze the data (Al-Moraissi et al., 2023). These functions enabled us to perform calculations, organize the data, and derive meaningful insights. Out of the initial 5,402 highly cited documents, we found 4,878 documents with a $TC_{2022}$ of 100 or more, accounting for 90% of the initial set. This subset of documents formed the basis for our analysis. Additionally, to refine our search strategy and ensure the inclusion of relevant documents, we applied a filter known as the "front page" approach (Wang and Ho, 2011; Al-Moraissi et al., 2023). This approach involved considering the title, abstract, and author keywords as a filter for the search keywords in the Web of Science Core Collection's Topic (TS) field. As a result, we identified 4,851 documents (99% of the 4,878 documents) with the search keywords present in their "front page," establishing them as highly cited ML research publications. Figure 1 shows the representation for searching the highly cited machine learning publications in SCI-EXPANDED.

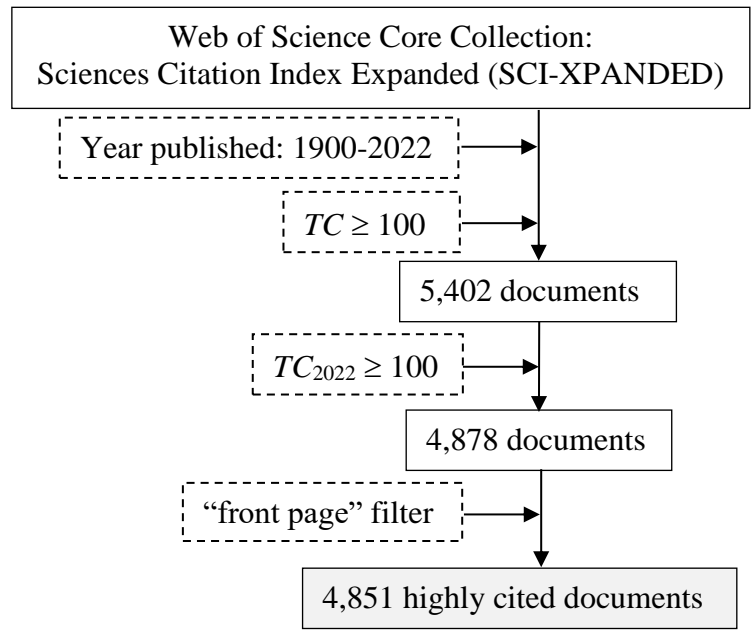

**Figure 1**. Schematic for searching the highly cited machine learning publications in SCI-EXPANDED

We obtained the journal impact factors ($IF_{2022}$) from the Journal Citation Reports (JCR) published in 2022 to assess the impact of the journals where these highly cited publications were published. This information further contributes to our analysis of the influence and prestige of the respective journals. By employing these rigorous methods and data analysis techniques, we aim to comprehensively and accurately examine highly cited ML research publications and their associated journal impact factors. These findings serve as a valuable resource for understanding the prominence and impact of ML research in the scholarly community.

**2.3 Authorship and affiliation handling**

In the SCI-EXPANDED database, the designation "reprint author" is used to identify the corresponding author. However, for this study, we adopted the term "corresponding author" to refer to the author with whom correspondence regarding the publication can be made (Chiu and Ho, 2007). It is important to note that in articles with unspecified authorship, single authors were considered both the first and corresponding authors (Ho, 2014b).

Similarly, for articles with unspecified corresponding institutions, the single institution listed was considered the first and corresponding-author institution (Ho, 2014b). In the case of articles from a single country, the country was classified as the first and corresponding-author country (Ho, 2014b). This approach ensures consistency and accuracy in assigning authorship and affiliations. For articles with multiple corresponding authors, institutions, and countries, all corresponding authors, their respective institutions, and countries were considered (Al-Moraissi et al., 2023). This allows for a comprehensive analysis that considers the contributions and affiliations of all relevant authors involved in the publication.

Additionally, a thorough verification process was conducted to address articles in the SCI-EXPANDED database where corresponding authors were listed with only addresses but not affiliation names. These articles were carefully examined, and the addresses were updated to include the corresponding affiliation names (Al-Moraissi et al., 2023). This step ensures that the affiliations of corresponding authors are accurately represented in our analysis. By employing these approaches to authorship and affiliation handling, we ensure that our analysis accurately represents the contributions of authors and their affiliated institutions, allowing for a comprehensive understanding of the research landscape in machine learning.

**2.4 Affiliation classification and publication performance evaluation**

In this study, we undertook a classification process for affiliations to ensure consistency and accuracy. Affiliations from England, Scotland, Northern Ireland, and Wales were reclassified as being from the United Kingdom (UK) (Chiu and Ho, 2005). This consolidation allows for a more comprehensive analysis of the contributions from the UK. Moreover, affiliations initially listed as Yugoslavia were carefully checked and reclassified as being from Slovenia (Wambu et al., 2017). This adjustment ensures the correct attribution of publications to their respective countries and facilitates accurate evaluation. To evaluate the publication performance of countries and institutions, we applied six publication indicators, as outlined by Hsu and Ho (2014):

- *TP* (Total Number of Articles): This indicator represents the total number of articles published by a specific country or institution.
- *IP* (Number of Single-Country Articles or Single-Institution Articles): *IP* denotes the number of articles where the authors are from a single country ($IP_C$) or a single institution ($IP_I$).
- *CP* (Number of Internationally Articles or Inter-Institutionally Collaborative Articles): *CP* signifies the number of articles resulting from international collaborations ($CP_C$) or inter-institutional collaborations ($CP_I$).
- *FP* (Number of First-Author Articles): *FP* refers to the number of articles where the authors are listed as the first authors.
- *RP* (Number of Corresponding-Author Articles): *RP* represents the number of articles where the authors are identified as the corresponding authors.
- *SP* (Number of Single-Author Articles): *SP* denotes the number of articles authored by a single author.

By utilizing these publication indicators, we gain valuable insights into the publication performance of countries and institutions, highlighting their level of collaboration, authorship patterns, and overall research output. This information aids in the assessment and comparison of their contributions to the field of machine learning.

## 2.5 Publication impact evaluation and *Y*-index

In addition to the six publication indicators, we also applied six citation indicators to evaluate the publication impact of countries and institutions (Ho and Mukul, 2021). One of the metrics used for evaluating the publication performance of authors is the *Y*-index. The *Y*-index, as defined by Ho (2012; 2014a), is denoted as the *Y*-index (*j*, *h*), where *j* is a constant related to the publication potential, determined by the sum of the first-author articles and corresponding-author articles. The parameter *h* is a constant associated with the publication characteristics and represents the polar angle indicating the proportion of corresponding-author articles to first-author articles.

The value of *j* indicates the contribution of the author as a first or corresponding author to the articles. A higher value of *j* suggests a greater contribution by the author in terms of first-author and corresponding-author articles. The parameter *h* is defined as follows:

- *h* = π/2: This indicates an author who has solely published corresponding-author articles (*j* represents the number of corresponding-author articles).
- π/2 > *h* > π/4: This range signifies an author who has a higher proportion of corresponding-author articles compared to first-author articles (*FP* > 0).
- *h* = π/4: This value indicates an author with an equal number of first-author and corresponding-author articles (*FP* > 0 and *RP* > 0).
- π/4 > *h* > 0: This range represents an author with more first-author articles than corresponding-author articles (*RP* > 0).
- *h* = 0: This value indicates an author who has exclusively published first-author articles (*j* represents the number of first-author articles).

By applying the *Y*-index, we can evaluate the publication performance of authors, taking into account their contribution as both first authors and corresponding authors. The *Y*-index provides insights into the author's publication potential and the proportion of their contributions based on the types of articles they have published. This metric enables a comprehensive assessment of authors' publication impact, considering their roles and contributions to the scholarly literature in machine learning.

## 3. Results and Discussion

In this section, we present the findings and engage in a discussion of our bibliometric analysis study, which is centered on publications in the field of machine learning and their citation patterns. Through this analysis, we aim to uncover and examine the most highly cited publications within the field of machine learning. Through an exploration of citation trends and patterns, we can gain valuable insights into the influential works that have shaped the landscape of ML research. Citation analysis provides a quantitative measure of the impact and significance of scholarly publications. Similarly, by investigating the citation counts, we can identify the publications that have gained substantial attention and recognition within the academic community. Furthermore, analyzing the citation patterns allows us to identify the key works that have contributed to the advancement of ML research and have had a lasting impact on the field.

The subsequent sections of this study provide a detailed description of our bibliometric analysis approach and the findings derived from the analysis of the collected data. We present an overview of the top cited ML publications and discuss the implications of these findings for the field. Additionally, we explore the characteristics of these highly cited works, such as their authors, publication journals, and prevalent themes. By comprehensively examining the most cited publications, we aim to identify the influential authors, notable research directions, and emerging trends within the field of machine learning. These insights will contribute to the existing body of knowledge and guide researchers, practitioners, and decision-makers in understanding the influential works and the evolving landscape of ML research. In the following sections, we present our findings and discuss the bibliometric analysis results comprehensively, shedding light on the key contributions and trends within the highly cited ML publications.

## 3.1. Characteristics of document types

The approach described by Monge-Nájera and Ho (2017) to regenerate the characteristics of a document type involves utilizing two key metrics: the average number of citations per publication ($CPP_{year}$) and the average number of authors per publication ($APP$). This approach has been employed in the bibliometric analysis of highly cited articles published in SCI-EXPANDED, as documented by Ho and Shekofteh (2021) and Ho and Ranasinghe (2022). To apply this approach, we followed these steps:

- Define the document type: Specify the particular document type you wish to analyze based on the available data. It could be scientific articles, research papers, or another relevant category.

- Gather the necessary data: Obtain a dataset that provides the total number of citations ($TC_{year}$), the total number of publications ($TP$), and the total number of authors ($AU$) for each document of the chosen type. Ensure the dataset covers the desired time frame and corresponds to articles published in SCI-EXPANDED.

- Calculate the average number of citations per publication ($CPP_{year}$): Divide the total number of citations in a given year ($TC_{year}$) by the total number of publications ($TP$) in that same year. This calculation will yield the average number of citations per publication for the specified document type. $CPP_{year} = TC_{year}/TP$

- Calculate the average number of authors per publication ($APP$): Divide the total number of authors ($AU$) by the total number of publications ($TP$) for the document type. This calculation will provide the average number of authors per publication. $APP = AU/TP$

- Analyze the results: Examine the obtained values of $CPP_{year}$ and $APP$ to gain insights into the characteristics of the document type. A higher $CPP_{year}$ suggests a greater average number of citations per publication, indicating a higher impact or visibility. Conversely, a higher *APP* signifies increased collaboration and multiple authorship within the document type.

- Contextualize with existing literature: Consider the findings of Monge-Nájera and Ho (2017), Ho and Shekofteh (2021), and Ho and Ranasinghe (2022) to compare and contextualize your results. This will enable validation of your regenerated characteristics and provide additional perspectives on the document type's attributes.

It is important to note that the specific methodologies and any additional factors accounted for in the aforementioned study steps may influence the precise results. Therefore, a thorough review and understanding of the methodologies employed in each related study are essential for meaningful comparisons and interpretations of any regenerated characteristics.

In the analysis of SCI-EXPANDED, a total of 4,851 documents were found to contain search keywords in their "font page." These documents represent seven different document types specified in Table 1. Among the identified documents, there were 4,139 articles, accounting for 85% of the total. The average number of authors per publication (*APP*) for these articles was 5.8. Within the document types, reviews constituted 674 documents and had the highest $CPP_{2022}$ value of 373. This high $CPP_{2022}$ value could be attributed to a specific review titled "Gradient-based learning applied to document recognition" (Lecun et al., 1998), which had a total citation count ($TC_{2022}$) of 24,111. Among the 160 classic publications with a $TC_{2022}$ of 1,000 or more (such as Long et al., 2014), 32 were categorized as reviews. Proceedings papers accounted for seven highly cited publications, followed by four editorial materials and one book chapter.

The $CPP_{2022}$ value for reviews was found to be 1.3 times higher than that of articles. It was also noted that highly cited medical-related documents, such as those on multiple sclerosis (Ho and Ranasinghe, 2022) and insulin resistance (Ho and Shekofteh, 2021), had lower $CPP_{2022}$ values of 1.1 and 0.85, respectively, compared to reviews. A total of 674 reviews were published across 382 journals. The journal "Expert Systems with Applications" had the highest reviews, totaling 15. Additionally, it was observed that some documents could be categorized under multiple document types in the Web of Science Core Collection. For example, 180 proceedings papers, eight data papers, seven book chapters, and two retracted publications were also classified as articles. Therefore, the cumulative percentages in Table 1 may exceed 100% (Usman and Ho, 2020).

Furthermore, it is important to acknowledge that contributions can differ among various types of documents. In the case of articles, they generally consist of sections like introduction, methods, results, discussion, and conclusion. Based on this consideration, articles were chosen for closer analysis, specifically focusing on 4,139 highly cited ML articles published in English. It is worth mentioning that among these articles, one publication appeared in a bilingual journal, presenting content in both English and Estonian languages. This bilingual approach allowed for wider dissemination of the research findings to both English-speaking and Estonian-speaking audiences.

**3.2. Characteristics of publication outputs**

The study conducted by Ho and Shekofteh (2021) applied a correlation analysis between the annual number of highly cited articles ($TP$) and their $CPP_{year}$ in the medical topic of multiple sclerosis. This analysis aimed to gain insights into the development trends and impacts of articles in this field. Figure 2 illustrates the distribution of these highly cited articles. Among the highly cited articles, a notable observation is that the highest number of articles, specifically 575, were published in 2018. Notably, it took a full five years for the number of articles to reach their peak, which was considerably longer than the highly cited multiple sclerosis articles, which took 12 years to reach their peak.

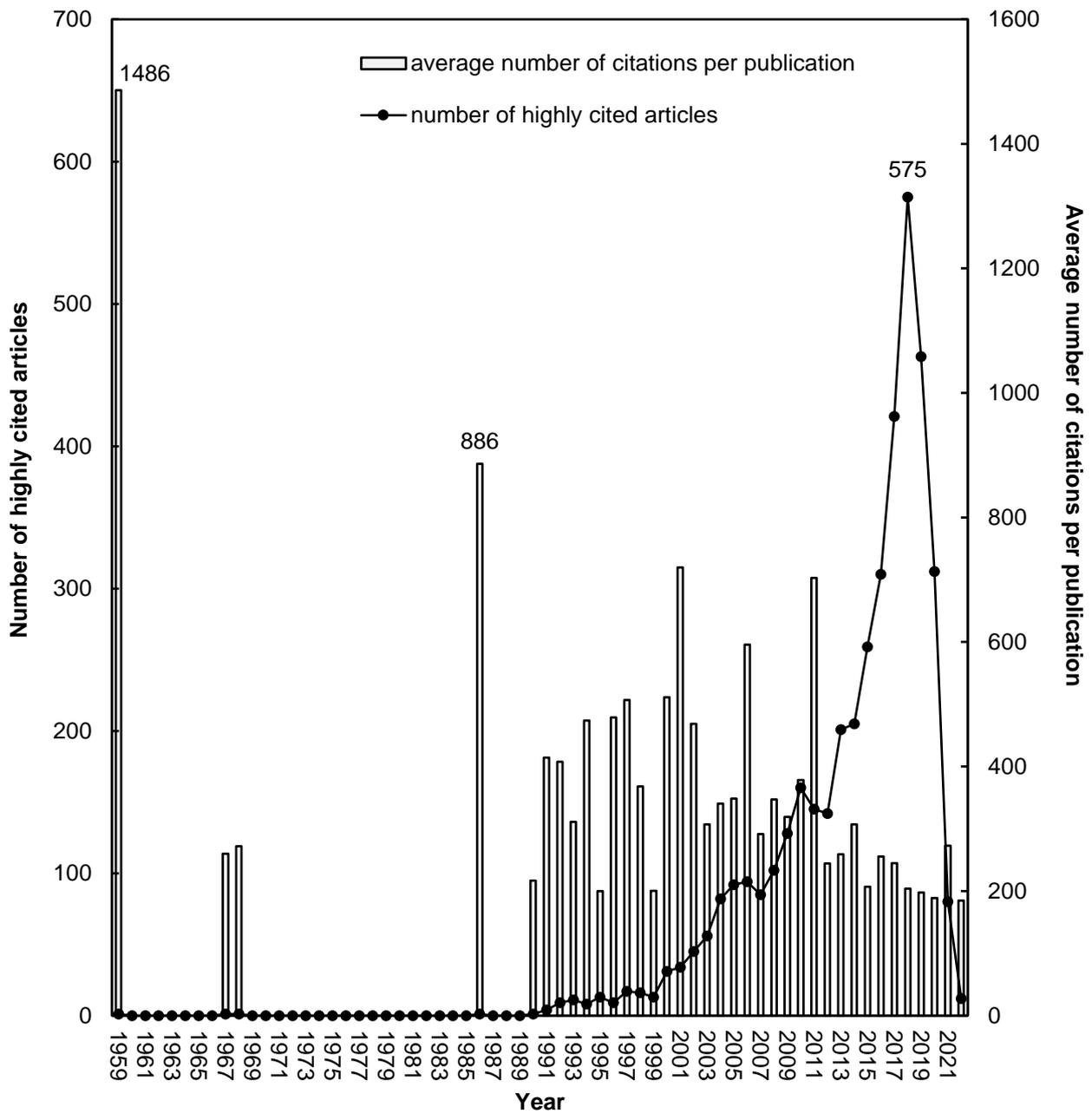

**Figure 2**. Number of highly cited machine learning articles and their average number of citations per publication by year.

The field of ML has seen increased activity from authors and researchers. In 1959, a significant article titled "Some studies in ML using the game of checkers" (Samuel, 1959) had the highest $CPP_{2022}$ value of 1,486. This suggests the influential impact of this article in the field of ML, which has gained prominence as a relatively new research area. These findings highlight the dynamics and growth patterns within the medical topic of multiple sclerosis and the role of ML in shaping the research landscape.

Similarly, in 1986, Farmer et al. introduced the article titled "The immune-system, adaptation, and machine learning," which garnered a high $CPP_{2022}$ score of 886. The paper investigates the relationship between the immune system, adaptation, and machine learning. It explores how principles derived from the immune system, known for its vital role in safeguarding the body against pathogens, can be applied to the realm of machine learning. The authors aimed to harness the immune system's learning, memory, and pattern recognition capabilities to construct a dynamic model based on Jerne's network hypothesis. By combining concepts from immunology and ML, the research offers the potential for developing adaptive learning algorithms inspired by biological systems. This interdisciplinary approach provides valuable insights into the synergy between the immune system and ML, with the goal of advancing the field.

### 3.3. Web of Science Category and Journal

In 2022, the Journal Citation Reports (JCR) included a total of 9,510 journals that contained citation references across 178 categories in the SCI-EXPANDED section of Web of Science. Within these categories, there were 1,007 journals that published highly cited articles specifically related to machine learning. These articles were distributed among 152 Web of Science categories within the SCI-EXPANDED section. The majority of the ML articles, constituting 22% of a total of 4,139 articles, were published in the field of artificial intelligence computer science, amounting to 907 articles. Furthermore, 794 articles (19%) were published in the field of electrical and electronic engineering, 459 articles (11%) in the field of information systems computer science, 429 articles (10%) in the field of interdisciplinary applications computer science, and 300 articles (7.2%) in the field of multidisciplinary sciences. It is evident that ML has gained significant attention and has been applied across a wide range of research fields.

A recent study conducted by Ho (2021) proposed utilizing the average number of citations per publication ($CPP_{year}$) and the average number of authors per publication ($APP$) as key indicators to characterize journals within a specific research topic. Table 2 in the study presents information on the top 11 most productive journals that published 40 or more highly cited articles. This includes their journal impact factors, $CPP_{2022}$, and $APP$. The Journal of ML Research, with an impact factor of 5.177 in 2021, published the highest number of highly cited articles, amounting to 88. These articles accounted for only 2.1% of the total 4,139 highly cited articles, indicating that ML is a relatively new research topic in various fields. Among the top 11 productive journals, the highly cited ML articles published in the Journal of ML Research achieved the highest $CPP_{2022}$, reaching 1,041. Interestingly, three of the top ten most cited articles, authored by Pedregosa et al. (2011), Srivastava et al. (2014), and Demsar (2006), were published in the Journal of ML Research. In contrast, the journal *Expert Systems with Applications* ($IF_{2022}$ = 8.5) had a significantly lower number of highly cited articles, with only 190. The average number of authors per publication ($APP$) varied across journals, ranging from 11 in Nature Communications to 3.1 in *Expert Systems with Applications*.

When considering the impact factor ($IF_{2022}$) of the top five journals with an $IF_{2022}$ more than 100, it was found that the *CA-A Cancer Journal for Clinicians* ($IF_{2022}$ = 254.7) had one article, the *Lancet* ($IF_{2022}$ = 168.9) had two articles, the *New England Journal of Medicine* ($IF_{2022}$ = 158.5) had two articles, the *JAMA-Journal of the American Medical Association* had five articles ($IF_{2022}$ = 158.5) had five articles, and the *BMJ-British Medical Journal* ($IF_{2022}$ = 105.7) had one article.

These five journals were among the highest ranked in their respective categories, with the *Lancet*, the *New England Journal of Medicine*, the *JAMA-Journal of the American Medical Association*, the *BMJ-British Medical Journal* securing the 1st, 2nd, 3rd, and 4th positions among 167 journals in the Web of Science category of general and internal medicine. Furthermore, the *CA-A Cancer Journal*

*for Clinicians* ranked first not only in the category of oncology (241 journals) but also in the SCI-EXPANDED (9,510 journals).

### 3.4. Publication performances: countries and institutions

The significant contributions of two authors, namely the first author and the corresponding author, in a research article have been widely acknowledged (Riesenberg and Lundberg, 1990). Within the SCI-EXPANDED dataset, seven highly cited ML articles (1.7% of the total 4,139 highly cited articles) did not have affiliations listed. On the other hand, a total of 4,132 highly cited articles were published by authors affiliated with 102 different countries. Among these, 2,460 articles (60% of 4,132) were single-country articles published by authors from 58 countries, with a $CPP_{2022}$ of 294. Additionally, 1,672 internationally collaborative articles (40%) were published by authors from 102 countries, with a $CPP_{2022}$ of 268. The results indicated that internationally collaborative research had a slightly lower citation impact in the highly cited ML domain.

To compare the productivity of different countries, six publication indicators and six related citation indicators ($CPP_{2022}$) were utilized (Ho and Mukul, 2021). Table 3 presents the findings for the top 15 productive countries, each having more than 100 highly cited articles. Notably, Egypt ranked 35th with 25 articles and emerged as the most productive country in Africa. Among the publication indicators, the USA led in all six categories: *TP* (1,932 highly cited articles, 47% of the total), $IP_C$ (1,045 articles, 42% of single-country articles), $CP_C$ (887 articles, 53% of internationally collaborative articles), *FP* (1,418 articles, 34% of first-author articles), *RP* (1,480 articles, 36% of corresponding-author articles), and *SP* (72 articles, 39% of single-author articles). When comparing the top 15 productive countries, France achieved the highest $CPP_{2022}$ in various categories: *TP* (453), *FP* (595), and *RP* (570) for ML research. Canada had the highest $CPP_{2022}$ of 526 for $IP_C$, while Japan had the highest $CPP_{2022}$ of 580 for $CP_C$. With four articles in the SP category, Australia attained the highest $CPP_{2022}$ of 1,138. These findings shed light on the productivity and citation impact of different countries in the field of highly cited ML research.

According to Ho (2012), the institution of the corresponding author in a research article often represents either the study's home base or the paper's origin. In terms of institutions, 1,288 highly cited ML articles (31% of the total 4,132 articles) were attributed to single institutions, with a $CPP_{2022}$ of 322. On the other hand, 2,844 articles (69%) were the result of institutional collaborations, with a $CPP_{2022}$ of 266. These findings suggest that institutional collaborations contribute to higher citation rates.

Table 4 provides an overview of the top 15 productive institutions and their respective characteristics. Among them, five institutions were located in the United States, three in the United Kingdom, two in China, and one each in Canada and Singapore. Notably, Zagazig University in Egypt, the University of KwaZulu-Natal in South Africa, and Cairo University in Egypt emerged as the most productive institutions in Africa, with each institution publishing six highly cited articles and ranking 360 in the dataset.

The Massachusetts Institute of Technology (MIT) in the United States demonstrated its dominance in four out of six publication indicators. It achieved a *TP* (Total Productivity) of 127 highly cited articles, accounting for 3.1% of the total 4,132 highly cited articles. Additionally, MIT had a $CP_I$ (Collaboration Productivity Index) of 102 articles, representing 3.6% of the 2,844 inter-institutionally collaborative articles. In terms of *FP* (First-author Productivity), MIT contributed 70 articles, making up 1.7% of the 4,132 first-author articles. Similarly, MIT had an *RP* (Corresponding-author Productivity) of 75 articles, which accounted for 1.8% of the 4,124 corresponding-author articles.

In contrast, Stanford University in the United States secured the top position in terms of the Institutional Productivity Index ($IP_I$) with 27 articles. This accounted for 2.1% of the total 1,288 single-institution articles. On the other hand, the University of Wisconsin, also in the United States, emerged as the leader in Single-author Productivity (*SP*) with 47 articles, representing 3.2% of the 186 single-author articles.

In comparison to the top 15 productive institutes listed in Table 4, the University of Washington in the United States demonstrated outstanding performance. It achieved a *TP* (Total Productivity) of 53 articles and a $CP_I$ (Collaboration Productivity Index) of 47 articles. Remarkably, the University of Washington had the highest $CPP_{2022}$ of 939 and 1,038 for *TP* and $CP_I$, respectively. However, the University of Toronto in Canada excelled in terms of the Institutional Productivity Index ($IP_I$), First-author Productivity (*FP*), and Corresponding-author Productivity (*RP*). With an $IP_I$ of six articles, an *FP* of 20 articles, and an *RP* of 23 articles, the University of Toronto boasted impressive $CPP_{2022}$ values of 3,494, 1,321, and 1,173 for $IP_I$, *FP*, and *RP*, respectively. Furthermore, the University of California, Berkeley in the United States, with three articles in the Single-author Productivity (*SP*) category, achieved an exceptional $CPP_{2022}$ of 4,804.

It is noteworthy that both MIT and Stanford University are renowned institutions known for their contributions to research, including early ML research. Several factors contribute to their reputation in this field:

- Strong Faculty: MIT and Stanford have attracted and cultivated world-class faculty members in various disciplines, including computer science and artificial intelligence. These faculties often conduct groundbreaking research and attract top talent, contributing to the overall research excellence of these institutions.
- Research Funding: Both institutions have a history of securing substantial research funding, which allows researchers to pursue ambitious projects and support their work. Adequate funding provides the resources needed for conducting experiments, accessing data, and developing innovative algorithms, giving them a competitive edge in producing impactful research.

- Collaborative Environment: MIT and Stanford foster a collaborative research environment, encouraging interdisciplinary collaboration among researchers, students, and industry partners. This promotes knowledge sharing, facilitates the exchange of ideas, and encourages cross-pollination of expertise from different fields, leading to innovative solutions and breakthroughs.

- Access to Industry: Being situated in close proximity to tech hubs like Silicon Valley (Stanford) and the Boston-Cambridge area (MIT), these universities have strong connections with industry leaders and startups. This proximity offers opportunities for collaborations, internships, and access to real-world datasets and challenges, enabling researchers to address practical problems and apply their findings to industry applications.

- Academic Reputation: MIT and Stanford have long-standing reputations for excellence in education and research. Their strong academic standing attracts top-tier students and researchers from around the world. The high-calibre talent pool and rigorous academic programs foster an environment conducive to producing impactful research outcomes.

These factors and a strong institutional commitment to research and innovation contribute to MIT and Stanford's success in early ML research and their ongoing prominence in the field.

**3.5. Publication performances: authors**

In the domain of highly cited articles related to ML, the average number of authors per publication (*APP*) was 5.8. The maximum number of authors in a single article was 346 (Abolfathi et al., 2018). Out of the 4,139 articles with available author information, the majority, accounting for 65% of the total, were published by groups of two to five authors. Specifically, there were 782 highly cited articles (19% of the total) written by groups of 3 authors, 685 articles (17%) by groups of 4 authors, 657 articles (16%) by groups of 2 authors, and 549 articles (13%) by groups of 5 authors.

Table 5 presents the top 15 productive authors who have contributed 15 or more highly cited ML articles. Among them, Muller emerged as the most productive author, with 35 highly cited articles, including one as the first author and 15 as the corresponding author. Zhou published 21 articles, out of which 12 were as the first author. Additionally, Onan published seven articles, with five of them being single-author articles. Comparing the 15 productive authors, Zhou achieved the highest $CPP_{2022}$ (citation per publication in 2021) values across all categories, with scores of 461 for all highly cited articles, 361 for first-author articles, and 491 for corresponding-author articles. Only three of the top 15 authors, Zhang, Von Lilienfeld, and Liu, had single-author articles. Notably, eight of the 15 productive authors, including Zhou, Pham, Zou, Zhang, Muller, Liu, Bui, and Von Lilienfeld, were recognized as top authors in terms of publication potential, as evaluated by the $Y$-index.

Out of the total 4,139 highly cited ML articles, a vast majority of 4,125 articles (99.7% of the total) included information about both the first author and corresponding author in SCI-EXPANDED. These articles were thoroughly analyzed using the $Y$-index as a metric. The 4,125 highly cited ML articles involved a total of 18,234 authors. Among them, 13,399 authors (73% of the total) had no first-author or corresponding-author articles, resulting in a $Y$-index value of (0, 0). Additionally, 1,365 authors (7.5%) exclusively published corresponding-author articles with a h-index value of $\pi/2$. Furthermore, 133 authors (0.73%) had more corresponding-author articles than first-author articles, with $\pi/2 > h > \pi/4$, indicating a higher h-index value for corresponding-author articles compared to first-author articles ($FP > 0$). Meanwhile, 2,001 authors (11%) contributed an equal number of first-author and corresponding-author articles, resulting in an h-index value of $\pi/4$ ($FP > 0$ and $RP > 0$). In contrast, 128 authors (0.70%) published more first-author articles than corresponding-author articles, with $\pi/4 > h > 0$, signifying a higher h-index value for first-author articles ($RP > 0$). Finally, 1,208 authors (6.6%) exclusively published first-author articles, yielding an $h$-index value of 0. These analyses were conducted to explore the authorship patterns and productivity of highly cited ML articles based on the *Y*-index.

The polar coordinate plot shown in Figure 3 illustrates the distribution of the *Y*-index ($j$, $h$) for the top 41 potential authors in highly cited ML research, where $j \geq 8$. Each point on the plot represents a coordinate *Y*-index ($j$, $h$) corresponding to a single author or multiple authors. For instance, authors Chen, Kononenko, Wang, Chicco, Yuan, Ishibuchi, Zhang, and Wang have a *Y*-index of ($8$, $\pi/4$), while Chen and Koutsouleris have a *Y*-index of ($9$, $0.6747$). Among these potential authors, Zhou has the highest publication potential in highly cited ML articles, with a *Y*-index of ($24$, $\pi/4$), followed by Pham, with a *Y*-index of ($20$, $0.8851$). Authors Ramprasad, Deo, Chen, and seven others share the same *j*-value of 8, indicating they have the same publication potential in highly cited ML research. However, they exhibit different publication characteristics. Ramprasad has published only eight corresponding-author articles with an *h*-value of $\pi/2$, while Deo has published more corresponding-author articles than first-author articles with an h-value of 1.030. Chen and the other seven authors have an equal number of first-author and corresponding-author articles, resulting in an h-value of $\pi/4$. Manavalan has published more first-author articles than corresponding-author articles, with an *h*-value of 0.5404.

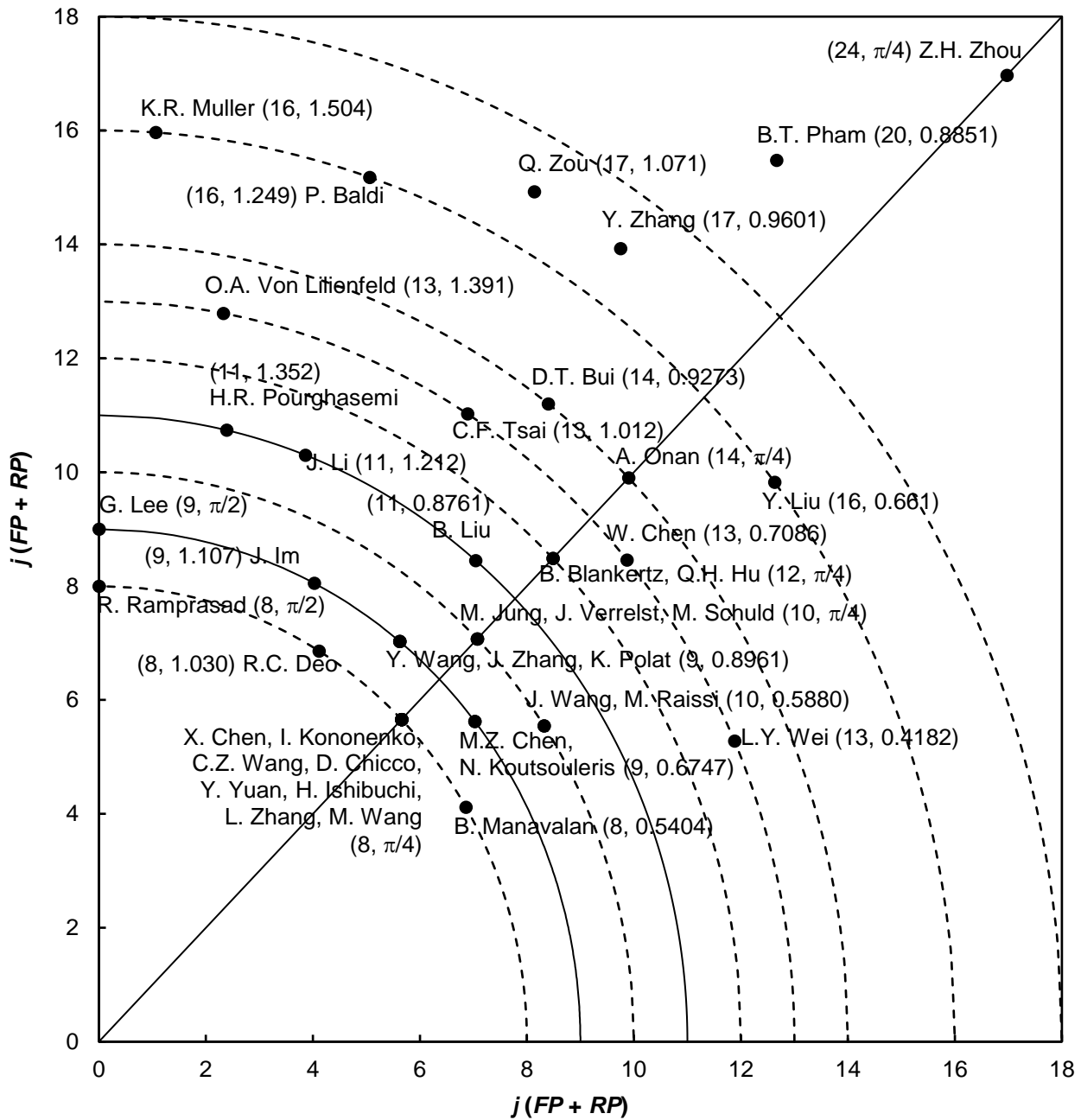

**Figure 3.** Top 41 authors with *Y*-index ($j \geq 8$)

Similar patterns were observed for authors with *j*-values of 9, 10, 11, 13, 14, 15, and 16. For example, authors Zhou (24, π/4), Onan (14, π/4), B. Blankertz (12, π/4), Hu (12, π/4), Jung (10, π/4), Verrelst (10, π/4), Schuld (10, π/4), and Zhang with seven other authors (8, π/4) are all located on the diagonal line representing $h = π/4$. This indicates that they share similar publication characteristics but differ in their publication potential. Zhou has the highest publication potential with a *j*-value of 24, followed by Onan with a *j*-value of 14, Blankertz and Hu with a *j*-value of 12, Jung, Verrelst, and Schuld with a $j$-value of 10, and Zhang with seven other authors with a $j$-value of 8. Similarly, authors Lee (9, $\pi/2$) and Ramprasad (8, $\pi/2$) are located on the y-axis representing $h = \pi/2$, indicating that they share the same publication characteristics. However, Lee has a greater publication potential compared to Ramprasad. It is important to note that the authorship analysis may be subject to potential biases due to authors with the same name or the same author using different names over time, which can impact the accuracy of the findings (Chiu and Ho, 2007).

**3.6. The top ten most frequently cited articles in machine learning research**

The total citations ($TC$) of articles are periodically updated in the Web of Science Core Collection. In order to enhance the accuracy of bibliometric studies, the total number of citations from the Web of Science Core Collection, specifically from the publication year until the end of 2022 ($TC_{2022}$), was utilized to minimize bias, as suggested by Wang et al. (2011). Among the analyzed articles, 908 articles (22% of 4,139 articles), 3,375 articles (82% of 4,117 articles with abstracts in SCI-EXPANDED), and 1,511 articles (51% of 2,984 articles with author keywords in SCI-EXPANDED) contained search keywords in their title, abstract, and author keywords, respectively.

Table 6 presents the top ten most frequently cited articles in ML research. Among these articles, one article had search keywords in its title, nine had search keywords in its abstracts, and two had search keywords in its author keywords. One of the most frequently cited and referred-to articles concerning the paper titled "Statistical Comparisons of Classifiers over Multiple Data Sets" is the work authored by Garcia and Herrera in 2008. In their publication, "An extension on 'Statistical Comparisons of Classifiers over Multiple Data Sets' for all pairwise comparisons," Garcia and Herrera introduced an extension to a previously established statistical framework used for the evaluation of the performance of multiple classifiers across various datasets. This article was featured in the Journal of ML Research, specifically in volume 9, encompassing pages 2677-2694.

Building upon the foundation laid by earlier research, the authors extended the scope of statistical comparisons to encompass pairwise evaluations among all classifiers. They also introduced novel statistical tests and procedures designed to provide a comprehensive assessment of performance variations among classifiers across a multitude of datasets. This extension significantly enhances the previous framework by delivering a more in-depth and detailed analysis of classifier performance.

The article contributes to the field of ML by addressing the need for rigorous statistical comparisons when evaluating and selecting classifiers. By considering all possible pairwise comparisons, the authors provide a comprehensive approach to assess the relative performance of classifiers on diverse datasets. This extension has practical implications for researchers and practitioners in ML, as it offers a robust methodology for making informed decisions about classifier selection and performance evaluation. Inclusively, Garcia and Herrera's study presented a valuable extension to the existing framework for comparing classifiers over multiple datasets, enhancing the statistical analysis and providing a more comprehensive understanding of classifier performance.

In the course of our bibliometric analysis study on ML, we discovered five influential and widely cited articles within the field. These articles, published in various years, have played a crucial role in advancing ML research and have significantly impacted the field. Although they may not be among the top ten most frequently cited, they have contributed substantially and shaped the machine learning landscape. Subsequently, the next subsections give brief details of these articles and explore their notable contributions:

***Scikit-learn:*** *Machine learning in Python (Pedregosa et al., 2011)*
This article has reaped considerable attention and citations since its publication in 2011. It introduced the scikit-learn library, a powerful and widely used ML toolkit in Python. The article presented a comprehensive overview of the library's capabilities, providing researchers and practitioners with a valuable resource for developing ML models and conducting experiments. Its impact lies in enabling widespread adoption and facilitating the development of ML algorithms and applications. The articles were published by 16 authors from ten institutions in France, Japan, Germany, USA, and the UK, with a $C_{2022}$ of 7,359 (rank 1st in highly cited ML research) and a $TC_{2022}$ of 29,958 (rank 1st). The article is not only the most frequently cited but also the most impactful in ML research. The article had a sharp increasing citation after its publication to reach the top of ML research.

*Dropout*: A Simple Way to Prevent Neural Networks from Overfitting (Srivastava et al., 2014)

The article was published by five authors from the University of Toronto in Canada with a of 3,587 (rank 4th) and a $TC_{2022}$ of 19,590 (rank 3rd). This article introduced the concept of dropout regularization in deep learning. Dropout is a technique that mitigates overfitting by randomly dropping units during training. The article demonstrated the effectiveness of dropout in improving generalization and preventing overfitting in deep neural networks. This contribution has become a cornerstone in machine learning, enabling the training of more robust and accurate deep-learning models.

*Random forests (Breiman, 2001)*

The article was published by Breiman from the University of California, Berkeley in the USA, with a $C_{2022}$ of 4,689 (rank 2nd) and a $TC_{2022}$ of 14,127 (rank 4th). Figure 4 shows an increasing citation trend after its publication, sharply increasing in recent years to reach the top second in ML research. Breiman's article introduced the concept of random forests, a powerful ensemble learning method. Random forests combine multiple decision trees to create a robust and accurate prediction model. This article significantly influenced the ML field by presenting a scalable and efficient approach for handling complex datasets. Random forests have since become a widely adopted and successful technique in various domains, showcasing their practical relevance and impact.

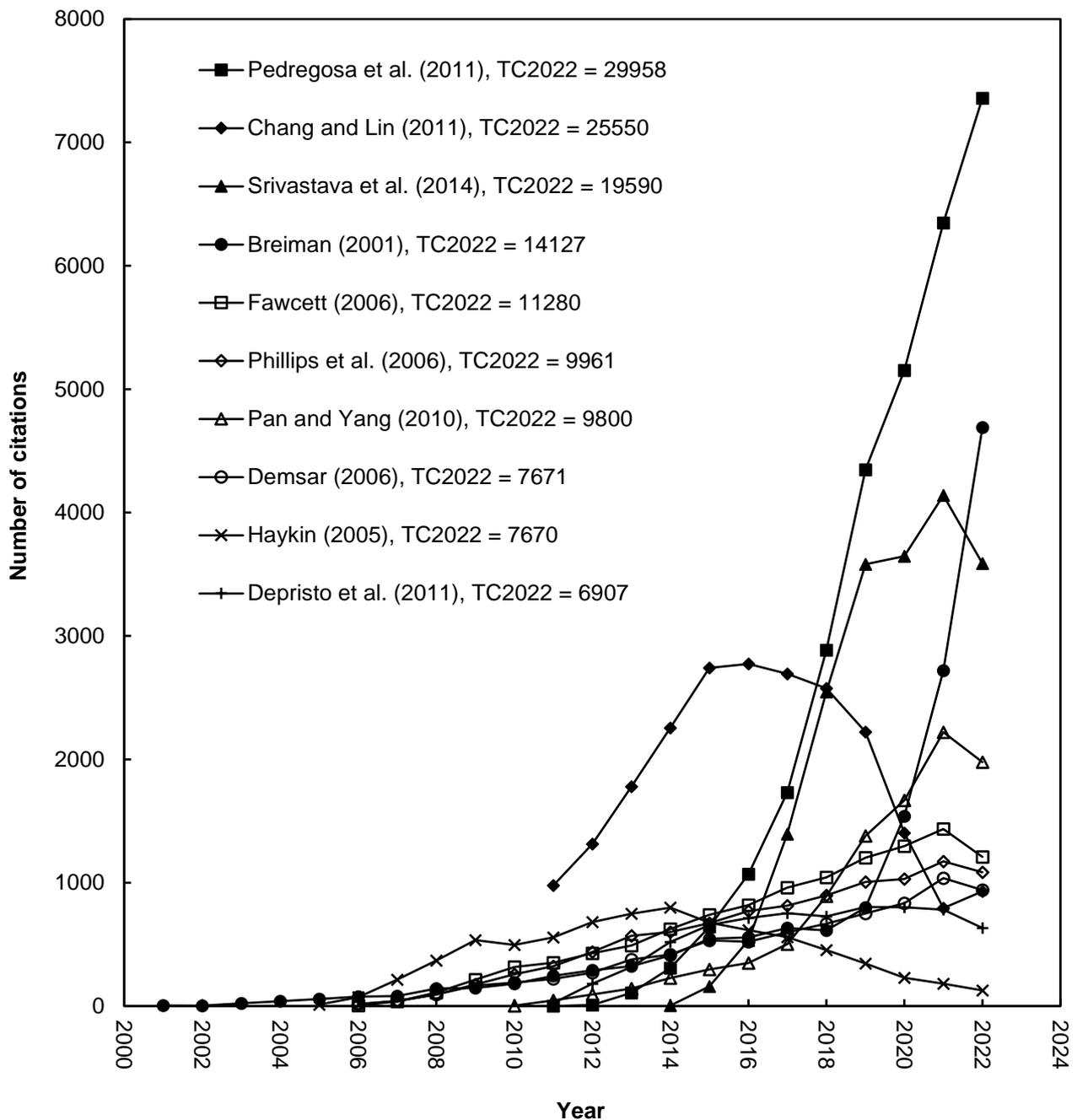

**Figure 4**. The citation histories of the top ten highly cited machine learning articles

*An introduction to ROC analysis (Fawcett, 2006)*

Fawcett's article introduced the concept of receiver operating characteristic (ROC) analysis, a widely used method for evaluating the performance of binary classifiers. The article provided a comprehensive overview of ROC analysis and its applications in ML. By introducing this evaluation technique, Fawcett's work has significantly influenced the field by enabling researchers to assess the performance of classifiers in a comprehensive and standardized manner. The article published by

Fawcett from the Institute for the Study of Learning and Expertise in the USA with a $C_{2022}$ of 1,210 (rank 8$^{th}$) and a $TC_{2022}$ of 11,280 (rank 5$^{th}$).

*A survey on transfer learning (Pan and Yang, 2010)*

Pan and Yang's article focused on the challenging problem of imbalanced learning, where the distribution of classes in the training data is heavily skewed. The article provided a comprehensive survey of techniques for addressing this issue, presenting various algorithms and strategies to tackle imbalanced learning problems. This work has been widely referenced and cited, providing researchers with a valuable resource for handling imbalanced datasets and advancing the field of ML. This article was published by authors from the Hong Kong University of Science and Technology in China with a $C_{2022}$ of 1,977 (rank 5$^{th}$) and a $TC_{2022}$ of 9,800 (rank 7$^{th}$).

Though not among the most frequently cited, these five articles have made significant contributions to the field of ML. Their impact lies in introducing innovative techniques, frameworks, and evaluation methods, which have paved the way for further research and advancements. Through their notable contributions, they have shaped the landscape of ML and continue to influence the work of researchers and practitioners in the field.

### 3.7. Strengths and limitations of the study

In our current investigation, we have taken several steps to mitigate potential limitations. Firstly, we have made efforts to minimize data bias by ensuring that our selection of publications is not skewed towards any particular source, publication type, or author. This safeguards the generalizability of our analysis results. Additionally, we have diligently addressed the issue of publication lag, ensuring that we consider recent publications to avoid any temporal bias. Notably, our dataset exclusively comprises publication extracts from the Clarivate Analytics Web of Science Core Collection, guaranteeing the utilization of high-quality publications and citations, thus avoiding the risk of omitting relevant highly cited and impactful ML publications due to limitations in data sources or incomplete records.

However, it's crucial to acknowledge that achieving a completely flawless study is a challenging task, and our work is not exempt from potential limitations. Some possible limitations include the challenge of dealing with citation self-selection, as similar studies may also be influenced by authors' tendencies to cite works that align with their own research, potentially resulting in an overestimation of certain influential papers. Furthermore, the evolving nature of the machine learning field, along with changes in terminology and subfields, may impact the identification of highly impactful papers across different decades. Interdisciplinary collaboration is a common aspect of ML research, making it challenging to carefully categorize papers within traditional boundaries, which can occasionally lead to misclassifications.

The assessment of impact and significance is inherently somewhat subjective, as different scholars may have varying criteria for defining groundbreaking research. Additionally, our analysis may not fully account for temporal biases in citation patterns, such as the natural tendency for older papers to accumulate more citations over time due to their longevity. Lastly, our study's focus on "highly cited" and "impactful" publications might inadvertently exclude important yet less-cited works that have exerted a significant influence on the field. We recognize that these limitations provide opportunities for future research, and we hope that our findings will inspire and inform further exploration in this evolving and interesting area of research.

## 4. Conclusion and Future Research Directions

In this bibliometric exploration, we embarked on a comprehensive analysis of highly cited and high impact ML publications. Through the employment of bibliometric analysis techniques, we sought to uncover key trends, influential authors, top journals, and significant themes within this thriving field.

Our analysis shed light on the landscape of ML research, providing valuable insights into its progress and direction. By examining citation patterns, co-authorship networks, and publication trends, we identified the most influential research papers and collaborations that have shaped the field of ML research. Moreover, we gained a deeper understanding of the key research areas, methodologies, and applications that have garnered substantial attention and citation within the research community and specifically the ML research enthusiasts.

The findings of our study contribute to the existing body of knowledge in ML research and its relevant applications to wide areas of interest or fields. They offer valuable insights for researchers, practitioners, and decision-makers in identifying seminal works and understanding the evolving landscape of the field. These insights can inform future research directions, foster collaborations, and guide the development of innovative approaches and methodologies within the field of ML.

While this bibliometric exploration has provided significant insights, several avenues for future research can further enhance our understanding of highly cited and high impact ML publications. Some potential directions for future research include:

- Fine-grained analysis: Conduct a more granular analysis of specific subfields within ML to uncover unique trends and influential publications within each domain.
- Citation analysis over time: Examining the temporal dynamics of citations to identify emerging research trends, changes in citation patterns, and the evolution of highly influential works in ML.
- Collaboration analysis: Investigating the collaborative networks and patterns of co-authorship within highly cited publications to identify key research groups and their contributions to the field.

- Cross-disciplinary analysis: Exploring the interdisciplinary nature of ML research by analyzing citations and collaborations across multiple domains, such as healthcare, finance, and natural language processing.
- Text mining and topic modeling: Applying text mining and topic modeling techniques to extract and analyze the main themes, research topics, and emerging concepts within highly cited ML publications.

By further pursuing these research directions, we can better understand the landscape of highly cited and high-impact ML publications. This knowledge will further facilitate advancements in the field and contribute to developing cutting-edge ML techniques and applications.

**Conflicts of interest/Competing interests** - Not applicable.

**Funding** – Not applicable.

**Ethics approval** – Not applicable.

Consent to participate – Not applicable.

**Consent for publication** – All authors agreed to the publication of the paper.

**Availability of data and material** - The datasets generated and analyzed during the current study are available from the corresponding author on reasonable request.

**Code availability** – Not applicable.

**Authors' contributions** – AEE, JG and YSH contributed equally in the initial manuscript writing. All authors contributed in editing and revising the final draft paper.

**Acknowledgement**

NA.

**Table 1**. Citations and authors according to the document type.

| Document type | TP | % | AU | APP | $TC_{2022}$ | $CPP_{2022}$ |
|---|---|---|---|---|---|---|
| Article | 4,139 | 85 | 23,983 | 5.8 | 1,175,116 | 284 |
| Review | 674 | 14 | 3,198 | 4.7 | 251,094 | 373 |
| Proceedings paper | 180 | 3.7 | 684 | 3.8 | 51,563 | 286 |
| Editorial material | 38 | 0.78 | 156 | 4.1 | 13,001 | 342 |
| Book chapter | 17 | 0.35 | 51 | 3.0 | 5,408 | 318 |
| Data paper | 8 | 0.16 | 53 | 6.6 | 1775 | 222 |
| Retracted publication | 2 | 0.041 | 6 | 3.0 | 237 | 119 |

*TP*: number of publications; *AU*: number of authors; *APP*: average number of authors per publication; $TC_{2022}$: the total number of citations from Web of Science Core Collection since publication year to the end of 2022; $CPP_{2022}$: average number of citations per publication ($TC_{2022}/TP$).

**Table 2.** The top 11 most productive journals with 40 highly cited articles or more.

| Journal | TP (%) | $IF_{2022}$ | APP | $CPP_{2022}$ | Web of Science category |
| --- | --- | --- | --- | --- | --- |
| Journal of Machine Learning Research | 88 (2.1) | 6.0 | 4.0 | 1,041 | automation and control systems artificial intelligence computer science |
| Expert Systems with Applications | 82 (2.0) | 8.5 | 3.1 | 190 | artificial intelligence computer science electrical and electronic engineering operations research and management science |
| Bioinformatics | 77 (1.9) | 5.8 | 4.3 | 230 | biochemical research methods biotechnology and applied microbiology interdisciplinary applications computer science mathematical and computational biology statistics and probability |
| IEEE Transactions on Pattern Analysis and Machine Intelligence | 76 (1.8) | 23.6 | 3.6 | 407 | artificial intelligence computer science electrical and electronic engineering |

| Journal | TP (%) | IF$_{2022}$ | APP | CPP$_{2022}$ | Category |
|---|---|---|---|---|---|
| Proceedings of the National Academy of Sciences of the United States of America | 68 (1.6) | 11.1 | 7.0 | 287 | multidisciplinary sciences |
| IEEE Access | 63 (1.5) | 3.9 | 4.9 | 233 | information systems computer science electrical and electronic engineering telecommunications |
| PLoS One | 54 (1.3) | 3.7 | 6.2 | 219 | multidisciplinary sciences |
| Nature Communications | 43 (1.0) | 16.6 | 10.9 | 218 | multidisciplinary sciences |
| Neurocomputing | 43 (1.0) | 6.0 | 3.7 | 206 | artificial intelligence computer science |
| Remote Sensing of Environment | 40 (1.0) | 13.5 | 5.9 | 199 | environmental sciences remote sensing imaging science and photographic technology |
| IEEE Transactions on Knowledge and Data Engineering | 40 (1.0) | 8.9 | 3.3 | 516 | artificial intelligence computer science information systems computer science electrical and electronic engineering |

*TP*: total number of highly cited articles; %: percentage of articles in all highly cited machine learning articles; *IF*$_{2022}$: journal impact factor in 2022; *APP*: average number of authors per article; *CPP*$_{2022}$: average number of citations per paper (*TC*$_{2022}$/*TP*).

**Table 3**. Top 15 productive countries with more than 100 highly cited articles.

| Country | TP | TP | | $IP_C$ | | $CP_C$ | | FP | | RP | | SP | |
|---|---|---|---|---|---|---|---|---|---|---|---|---|---|
| | | R (%) | $CPP_{2022}$ | R (%) | $CPP_{2022}$ | R (%) | $CPP_{2022}$ | R (%) | $CPP_{2022}$ | R (%) | $CPP_{2022}$ | R (%) | $CPP_{2022}$ |
| USA | 1,932 | 1 (47) | 311 | 1 (42) | 320 | 1 (53) | 300 | 1 (34) | 307 | 1 (36) | 303 | 1 (39) | 711 |
| China | 753 | 2 (18) | 221 | 2 (12) | 248 | 2 (27) | 202 | 2 (14) | 225 | 2 (14) | 225 | 5 (5.4) | 235 |
| UK | 549 | 3 (13) | 311 | 3 (6.3) | 251 | 3 (24) | 335 | 3 (7.1) | 265 | 3 (7.6) | 260 | 3 (6.5) | 256 |
| Germany | 402 | 4 (10) | 356 | 4 (4.2) | 279 | 4 (18) | 383 | 4 (5.7) | 322 | 4 (5.9) | 317 | 2 (7.0) | 357 |
| Canada | 301 | 5 (7.3) | 347 | 5 (3.9) | 526 | 5 (12) | 262 | 5 (3.8) | 433 | 5 (4.0) | 419 | 3 (6.5) | 861 |
| Australia | 245 | 6 (5.9) | 269 | 7 (2.1) | 269 | 6 (12) | 269 | 6 (2.8) | 294 | 6 (3.3) | 288 | 8 (2.2) | 1138 |
| France | 190 | 7 (4.6) | 453 | 15 (1.5) | 180 | 7 (9.2) | 517 | 11 (2.0) | 595 | 10 (2.1) | 570 | 12 (1.6) | 139 |
| Italy | 174 | 8 (4.2) | 231 | 11 (1.7) | 192 | 8 (7.9) | 244 | 10 (2.1) | 226 | 11 (2.0) | 227 | 8 (2.2) | 131 |
| Switzerland | 173 | 9 (4.2) | 276 | 13 (1.7) | 253 | 8 (7.9) | 284 | 7 (2.2) | 241 | 8 (2.3) | 241 | 12 (1.6) | 351 |
| South Korea | 158 | 10 (3.8) | 236 | 7 (2.1) | 163 | 11 (6.3) | 272 | 12 (1.9) | 170 | 7 (2.5) | 176 | 17 (1.1) | 119 |
| Spain | 153 | 11 (3.7) | 265 | 9 (2.0) | 247 | 12 (6.2) | 274 | 8 (2.2) | 266 | 9 (2.2) | 255 | 12 (1.6) | 141 |
| Netherlands | 146 | 12 (3.5) | 309 | 18 (1.0) | 218 | 10 (7.2) | 328 | 14 (1.5) | 256 | 14 (1.6) | 258 | 23 (0.54) | 123 |
| India | 138 | 13 (3.3) | 210 | 6 (2.2) | 201 | 14 (5.0) | 215 | 8 (2.2) | 197 | 12 (2.0) | 201 | 23 (0.54) | 116 |
| Singapore | 126 | 14 (3.0) | 228 | 16 (1.3) | 212 | 13 (5.6) | 234 | 13 (1.8) | 248 | 13 (1.8) | 243 | 17 (1.1) | 506 |

| Japan | 126 | 14 (3.0) | 442 | 10 (1.8) | 186 | 15 (4.9) | 580 | 15 (1.5) | 200 | 14 (1.6) | 186 | 12 (1.6) | 122 |

*TP*: number of total highly cited articles; *TP R* (%): total number of articles and the percentage of total articles; *IP*$_C$ *R* (%): rank and percentage of single-country articles in all single-country articles; *CP*$_C$ *R* (%): rank and percentage of internationally collaborative articles in all internationally collaborative articles; *FP R* (%): rank and the percentage of first-author articles in all first-author articles; *RP R* (%): rank and the percentage of corresponding-author articles in all corresponding-author articles; *SP R* (%): rank and the percentage of first-author articles in all first-author articles; *CPP*$_{2022}$: average number of citations per publication (*CPP*$_{2022}$ = *TC*$_{2022}$/*TP*); N/A: not available.

**Table 4**. Top 15 productive institutions.

| Institution | TP | TP R (%) | CPP | IP$_I$ R (%) | CPP | CP$_I$ R (%) | CPP | FP R (%) | CPP | RP R (%) | CPP | SP R (%) | CPP |
|---|---|---|---|---|---|---|---|---|---|---|---|---|---|
| MIT, USA | 127 | 1 (3.1) | 335 | 2 (1.9) | 269 | 1 (3.6) | 352 | 1 (1.7) | 313 | 1 (1.8) | 302 | 25 (0.54) | 188 |
| Stanford Univ, USA | 126 | 2 (3.0) | 398 | 1 (2.1) | 338 | 2 (3.5) | 414 | 2 (1.6) | 410 | 2 (1.7) | 402 | 3 (1.6) | 855 |
| Chinese Acad Sci, China | 107 | 3 (2.6) | 202 | 10 (0.85) | 151 | 3 (3.4) | 208 | 5 (0.92) | 203 | 3 (1.3) | 193 | 25 (0.54) | 134 |
| Harvard Univ, USA | 85 | 4 (2.1) | 397 | 20 (0.54) | 215 | 4 (2.7) | 413 | 8 (0.77) | 276 | 7 (0.87) | 279 | 25 (0.54) | 125 |
| Univ Calif Berkeley, USA | 77 | 5 (1.9) | 470 | 4 (1.2) | 1130 | 6 (2.1) | 297 | 4 (1.0) | 616 | 4 (1.1) | 564 | 3 (1.6) | 4804 |
| Univ Oxford, UK | 73 | 6 (1.8) | 320 | 35 (0.39) | 208 | 5 (2.4) | 328 | 12 (0.56) | 213 | 12 (0.61) | 207 | 25 (0.54) | 389 |
| Univ Calif San Diego, USA | 71 | 7 (1.7) | 254 | 5 (1.1) | 274 | 8 (2.0) | 250 | 6 (0.87) | 243 | 6 (1.0) | 244 | N/A | N/A |
| Nanyang Technol Univ, Singapore | 63 | 8 (1.5) | 228 | 10 (0.85) | 233 | 10 (1.8) | 226 | 7 (0.82) | 218 | 8 (0.85) | 214 | 7 (1.1) | 506 |
| Carnegie Mellon Univ, USA | 63 | 8 (1.5) | 280 | 3 (1.8) | 199 | 18 (1.4) | 326 | 3 (1.0) | 258 | 5 (1.1) | 261 | 2 (2.7) | 226 |
| Univ Cambridge, UK | 62 | 10 (1.5) | 278 | 84 (0.23) | 908 | 7 (2.1) | 246 | 17 (0.46) | 383 | 14 (0.56) | 333 | N/A | N/A |
| UCL, UK | 60 | 11 (1.5) | 200 | 55 (0.31) | 249 | 9 (2.0) | 196 | 13 (0.53) | 176 | 14 (0.56) | 180 | N/A | N/A |
| Univ Penn, USA | 57 | 12 (1.4) | 233 | 15 (0.70) | 166 | 12 (1.7) | 246 | 11 (0.65) | 194 | 10 (0.73) | 251 | N/A | N/A |
| Univ Toronto, Canada | 57 | 12 (1.4) | 650 | 27 (0.47) | 3494 | 11 (1.8) | 316 | 15 (0.48) | 1321 | 14 (0.56) | 1173 | 25 (0.54) | 228 |

| Univ Washington, USA | 53 | 14 (1.3) | 939 | 27 (0.47) | 164 | 13 (1.7) | 1038 | 31 (0.36) | 352 | 27 (0.44) | 459 | N/A | N/A |
| Tsinghua Univ, China | 53 | 14 (1.3) | 202 | 12 (0.78) | 210 | 15 (1.5) | 200 | 8 (0.77) | 192 | 8 (0.85) | 191 | N/A | N/A |

*TP*: total number of highly cited articles; *TP R* (%): total number of articles and percentage of total articles; *IP$_I$ R* (%): rank and percentage of single-institute articles in all single-institute articles; *CP$_I$ R* (%): rank and percentage of inter-institutionally collaborative articles in all inter-institutionally collaborative articles; *FP R* (%): rank and percentage of first-author articles in all first-author articles; *RP R* (%): rank and percentage of corresponding-author articles in all corresponding-author articles; *SP R* (%): rank and percentage of single-author articles in all single-author articles; *CPP*: average number of citations per publication ($CPP_{2022} = TC_{2022}/TP$); N/A: not available.

**Table 5**. Top 15 productive authors with 15 highly cited articles or more

| Author | TP | | FP | | RP | | SP | | h | rank (j) |
|---|---|---|---|---|---|---|---|---|---|---|
| | rank (TP) | $CPP_{2022}$ | rank (FP) | $CPP_{2022}$ | rank (RP) | $CPP_{2022}$ | rank (SP) | $CPP_{2022}$ | | |
| K.R. Muller | 1 (35) | 430 | 457 (1) | 299 | 1 (15) | 319 | N/A | N/A | 1.504 | 5 (16) |
| Y. Zhang | 2 (32) | 196 | 5 (7) | 189 | 7 (10) | 192 | 13 (1) | 321 | 0.9601 | 3 (17) |
| Y. Liu | 3 (25) | 355 | 2 (9) | 168 | 14 (7) | 155 | 2 (2) | 110 | 0.6610 | 5 (16) |
| J. Li | 3 (25) | 281 | 43 (3) | 231 | 10 (8) | 201 | N/A | N/A | 1.212 | 16 (11) |
| J. Wang | 5 (23) | 193 | 8 (6) | 141 | 40 (4) | 127 | N/A | N/A | 0.5880 | 19 (10) |
| D.T. Bui | 6 (22) | 191 | 8 (6) | 245 | 10 (8) | 202 | N/A | N/A | 0.9273 | 8 (14) |
| Y. Wang | 6 (22) | 166 | 23 (4) | 184 | 25 (5) | 168 | N/A | N/A | 0.8961 | 24 (9) |
| Z.H. Zhou | 8 (21) | 461 | 1 (12) | 361 | 2 (12) | 491 | N/A | N/A | 0.7854 | 1 (24) |
| Y. Li | 9 (20) | 186 | 457 (1) | 180 | 84 (3) | 159 | N/A | N/A | 1.249 | 148 (4) |
| O.A. Von Lilienfeld | 10 (19) | 331 | 110 (2) | 140 | 4 (11) | 238 | 13 (1) | 131 | 1.391 | 10 (13) |
| Q. Zou | 10 (19) | 165 | 8 (6) | 205 | 4 (11) | 172 | N/A | N/A | 1.071 | 3 (17) |
| J. Zhang | 12 (18) | 182 | 23 (4) | 225 | 25 (5) | 178 | N/A | N/A | 0.8961 | 24 (9) |
| B.T. Pham | 12 (18) | 179 | 2 (9) | 165 | 4 (11) | 207 | N/A | N/A | 0.8851 | 2 (20) |
| L. Zhang | 14 (15) | 175 | 23 (4) | 207 | 40 (4) | 198 | N/A | N/A | 0.7854 | 31 (8) |

| H. Shahabi | 14 (15) | 158 | 457 (1) | 110 | 84 (3) | 116 | N/A | N/A | 1.249 | 148 (4) |

*TP*: total number of articles; *FP*: first-author articles; *RP*: corresponding-author articles; *SP*: single-author articles; *CPP*$_{2022}$: average number of citations per publication (*CPP*$_{2022}$ = *TC*$_{2022}$/*TP*); *j*: a *Y*-index constant related to the publication potential; *h*: a *Y*-index constant related to the publication characteristics; N/A: not available.

**Table 6**. The ten most frequently cited articles in machine learning research.

| Rank ($TC_{2022}$) | Rank ($C_{2022}$) | Title | Country | Reference |
|---|---|---|---|---|
| 1 (29,958) | 1 (7,359) | Scikit-learn: Machine learning in Python | France, Japan, Germany, USA, UK | Pedregosa et al. (2011) |
| 2 (25,550) | 13 (930) | LIBSVM: A library for support vector machines | Taiwan | Chang and Lin (2011) |
| 3 (19,590) | 4 (3,587) | Dropout: A simple way to prevent neural networks from overfitting | Canada | Srivastava et al. (2014) |
| 4 (14,127) | 2 (4,689) | Random forests | USA | Breiman (2001) |
| 5 (11,280) | 8 (1,210) | An introduction to ROC analysis | USA | Fawcett (2006) |
| 6 (9,961) | 11 (1,085) | Maximum entropy modeling of species geographic distributions | USA | Phillips et al. (2006) |
| 7 (9,800) | 5 (1,977) | A survey on transfer learning | China | Pan and Yang (2010) |
| 8 (7,671) | 12 (941) | Statistical comparisons of classifiers over multiple data sets | Slovenia | Demšar (2006) |
| 9 (7,670) | 276 (126) | Cognitive radio: Brain-empowered wireless communications | Canada | Haykin (2005) |
| 10 (6,907) | 24 (632) | A framework for variation discovery and genotyping using next-generation DNA sequencing data | USA | Depristo et al. (2011) |

$TC_{2022}$: the total number of citations from Web of Science Core Collection since publication year to the end of 2022; $C_{2022}$: number of citations of an article in 2022 only.